\documentclass[twocolumn,reprint,
 superscriptaddress, amsmath,amssymb, aps, prl]{revtex4-1}
\usepackage{graphicx}
\usepackage{amsmath}
\usepackage{color}
\usepackage{dcolumn}
\usepackage{bm}
\usepackage[
   bookmarks=true,
   colorlinks=true,
   hyperindex=true,
   citecolor=blue,
   linkcolor=blue,
   urlcolor=blue
]{hyperref}

\begin{document}
\title{Mitigating noise of residual electric fields for single Rydberg atoms with electron photodesorption}
\author{Bahtiyar Mamat}
\affiliation{State Key Laboratory of Magnetic Resonance and Atomic and Molecular Physics, Innovation Academy for
Precision Measurement Science and Technology, Chinese Academy of Sciences, Wuhan 430071, China}
\affiliation{School of Physics, University of Chinese Academy of Sciences, Beijing 100049, China}
\author{Cheng Sheng}
\email{shengcheng@wipm.ac.cn}
\affiliation{State Key Laboratory of Magnetic Resonance and Atomic and Molecular Physics, Innovation Academy for
Precision Measurement Science and Technology, Chinese Academy of Sciences, Wuhan 430071, China}
\author{Xiaodong He}
\email{hexd@wipm.ac.cn}
\affiliation{State Key Laboratory of Magnetic Resonance and Atomic and Molecular Physics, Innovation Academy for
Precision Measurement Science and Technology, Chinese Academy of Sciences, Wuhan 430071, China}
\affiliation{Wuhan Institute of Quantum Technology, Wuhan 430206, China}
\author{Jiayi Hou}
\affiliation{State Key Laboratory of Magnetic Resonance and Atomic and Molecular Physics, Innovation Academy for
Precision Measurement Science and Technology, Chinese Academy of Sciences, Wuhan 430071, China}
\affiliation{School of Physics, University of Chinese Academy of Sciences, Beijing 100049, China}
\author{Peng Xu}
\affiliation{State Key Laboratory of Magnetic Resonance and Atomic and Molecular Physics, Innovation Academy for
Precision Measurement Science and Technology, Chinese Academy of Sciences, Wuhan 430071, China}
\affiliation{Wuhan Institute of Quantum Technology, Wuhan 430206, China}
\author{Kunpeng Wang}
\affiliation{State Key Laboratory of Magnetic Resonance and Atomic and Molecular Physics, Innovation Academy for
Precision Measurement Science and Technology, Chinese Academy of Sciences, Wuhan 430071, China}
\author{Jun Zhuang}
\affiliation{State Key Laboratory of Magnetic Resonance and Atomic and Molecular Physics, Innovation Academy for
Precision Measurement Science and Technology, Chinese Academy of Sciences, Wuhan 430071, China}
\affiliation{School of Physics, University of Chinese Academy of Sciences, Beijing 100049, China}
\author{Mingrui Wei}
\affiliation{State Key Laboratory of Magnetic Resonance and Atomic and Molecular Physics, Innovation Academy for
Precision Measurement Science and Technology, Chinese Academy of Sciences, Wuhan 430071, China}
\affiliation{School of Physics, University of Chinese Academy of Sciences, Beijing 100049, China}
\author{Min Liu}
\affiliation{State Key Laboratory of Magnetic Resonance and Atomic and Molecular Physics, Innovation Academy for
Precision Measurement Science and Technology, Chinese Academy of Sciences, Wuhan 430071, China}
\author{Jin Wang}
\affiliation{State Key Laboratory of Magnetic Resonance and Atomic and Molecular Physics, Innovation Academy for
Precision Measurement Science and Technology, Chinese Academy of Sciences, Wuhan 430071, China}
\affiliation{Wuhan Institute of Quantum Technology, Wuhan 430206, China}
\author{Mingsheng Zhan}
\affiliation{State Key Laboratory of Magnetic Resonance and Atomic and Molecular Physics, Innovation Academy for
Precision Measurement Science and Technology, Chinese Academy of Sciences, Wuhan 430071, China}
\affiliation{Wuhan Institute of Quantum Technology, Wuhan 430206, China}
\date{\today}

\begin{abstract}

Rydberg atoms as versatile tools for quantum applications are extremely sensitive to electric fields. When utilizing these atoms, it becomes imperative to comprehensively characterize and mitigate any residual electric fields present in the environment. For single Rydberg atoms in optical tweezer arrays, we reveal that one of the background electric fields originates from electrons bound to the vacuum cell surface. These electrons are generated by the 297-nm light used for single-photon Rydberg excitation. Once the electrons are desorbed from the surface through exposure to ultraviolet light, the coherence of ground-Rydberg transition is enhanced significantly. Furthermore, we demonstrate collective oscillations in the full Rydberg blockade regime of four atoms. Our investigations will advance the control and reliability of Rydberg atom-based systems for various quantum technologies.

\end{abstract}
\pacs{03.67.-a,03,67.Lx,42.50.Dv,42.50.Ct}


\maketitle


\maketitle

\section{introduction}


The rapid development of single Rydberg atoms in optical tweezers, as a tunable physical system at the quantum level, is aimed at realizing quantum simulation~\cite{Bernien2017,Kim2018,Semeghini2021,Chen2023,Browaeys2020,Morgado2021}, quantum computation~\cite{Bluvstein2022,Graham2022,Evered2023} and other realms of quantum information processing~\cite{Saffman2010,Omran2019,Ebadi2022,Kim2022}. Since the large polarizability of Rydberg states with high principal quantum numbers results in extreme sensitivity to electric field environments, Rydberg atoms have not only been utilized to measure local electric fields in proximity to an atom-chip surface~\cite{Carter2012,Davtyan2018}, but bias electric fields controlled by electrodes can also be employed to tune the interaction between Rydberg atoms~\cite{Ravets2014}.

Coherent excitation~\cite{Miro2010,Kubler2010,Avigliano2014}, interaction~\cite{Bernien2017,Zeiher2017}, and entanglement~\cite{Jau2016,Picken2018,Omran2019,Schine2022,Graham2022} of Rydberg atoms represent substantial foundations for quantum science and technologies. However, for single atoms optically trapped in tweezers, Rydberg atoms often experience decoherence due to uncertainties of background electric fields generated by adsorbate dipoles and electrons on the surfaces of the vacuum cell~\cite{Miro2010,Abel2011,Hattermann2012,Hankin2014,Sedlacek2016}. On one hand, as they create spatially dependent gradient electric fields for single atoms, the instability of the atom position introduces detunings in the ground-Rydberg transition frequency~\cite{Ocola2022}. On the other hand, not only thermal fluctuations of adsorbate dipole moments introduce electric field noise~\cite{Naini2011}, but also changes in distribution of adsorbates and electrons result in long-term drifts of the transition frequency. Therefore, to ensure the reliable and precise manipulation of Rydberg atoms, it becomes crucial to thoroughly investigate and mitigate residual electric fields in experiment.

In previous investigations, a quartz surface with adsorption of Rb atoms can be transformed into a negative electron affinity (NEA) surface where electrons can be bound, and the total background electric fields results from a sum of the electric fields from both the adsorbates and electrons~\cite{Sedlacek2016}. However, intermittent jumps in the Rydberg energy levels caused by electric fields have been observed in two-photon Rydberg experiments~\cite{Graham2022,Ocola2022} and the explanation of this phenomenon remains unclear. Therefore, a more comprehensive understanding and careful characterization of these uncontrollable and undesirable electric fields caused by adsorbates and electrons is imperative, particularly as system feature sizes shrink.

Compared with two-photon Rydberg experiments, single-photon Rydberg dressing schemes where interactions are optically admixed to the ground state provide switchable light-induced interactions between ground-state atom, extended interaction ranges and increased effective lifetime~\cite{Hankin2014,Jau2016,Zeiher2017,Borish2020}. Here, we use 297-nm light to excite single $^{87}$Rb atoms trapped in optical tweezers to Rydberg state in a fused quartz cell. We find that one of the major sources of the background electric fields is from electrons on the cell surface, which are generated by the 297-nm light. To provide evidence, we introduce an additional off-resonant 297-nm light to modulate the electric field. Consequently, we observe changes in the DC electric field as the pulse duration of the 297-nm light is extended. To address this issue, we employ a method to desorb the electrons from the surface by exposing it to ultraviolet (UV) light. The coherence of ground-Rydberg transition is significantly enhanced since the noise of background electric fields mitigated. Moreover, we demonstrate collective oscillations in the full Rydberg blockade regime of four atoms. Our results will be important for quantum simulation based on Rydberg-dressed atom arrays and high-fidelity quantum gates with Rydberg atoms in demand for pure electric field environment.


\section{experimental apparatus for rydberg atoms}

\begin{figure}[htbp]
\centering
\includegraphics[width=8.6cm]{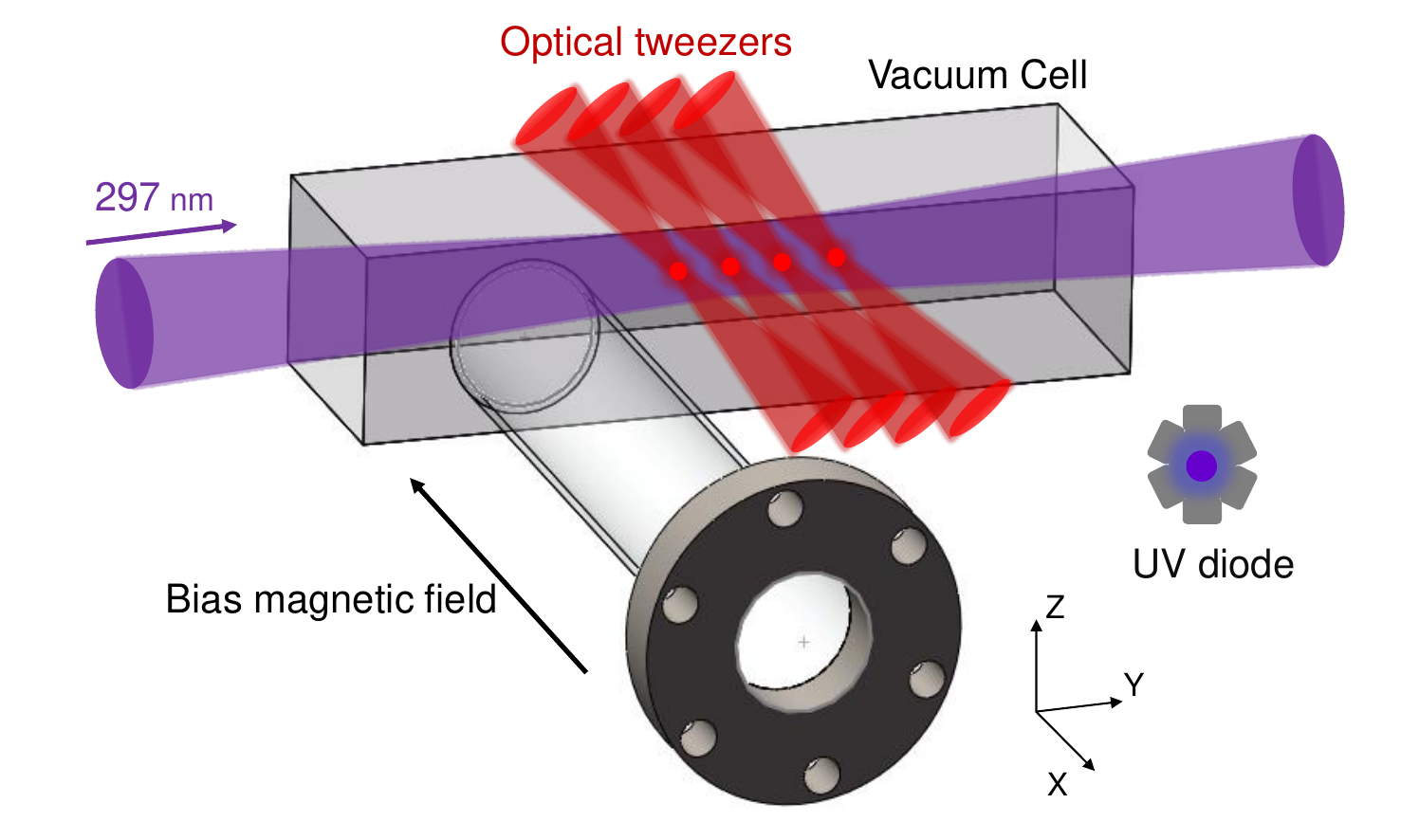}
\caption{(color online). Schematic of the optical layout for single-photon Rydberg excitation. Six UV diodes (TY-UV LED 365) with 3.43 W total power are used to homegeneously shine on the vaccum cell.}
\label{fig:fig1}
\end{figure}

Single $^{87}$Rb atoms are loaded into a $6\times4$ tweezer array from a magneto-optical trap in the center of the cella quartz vaccum cell (optical adhesive bonding and no coating, $34\times44\times134$ mm with 4 mm thickness, made by Japan Cell). Each 824-nm optical tweezer has a $1/e^{2}$ radius of 1 $\mu$m with trap depth of $U_{0}/k_{B}\approx$ 1 mK, where $k_{B}$ is the Boltzmann constant. Only one site in the array is then determined prepared via atom rearrangement. After the single atom optically pumped to $|5S_{1/2}, F=1, m_{F}=0\rangle$ state~\cite{Sheng2018}, we apply a 297-nm light pulse to excite it to Rydberg state 53$P_{3/2}$. A Rydberg state atom will experience an antitrapping potential leading to a quick ejection of the atoms from the optical dipole trap. So the probabilities of the Rydberg excitation are acquired by the atom loss rates. We check the atom loss via probing the atom fluorescence whose images are taken by an electron multiplying charge coupled device (EMCCD) camera. Our experimental setup is shown schematically in Fig.~\ref{fig:fig1} and the details of the 297-nm laser system can be seen in Appendix~\ref{appendix:A}A.


\section{rydberg spectroscopy and coherent excitation}

\begin{figure}[htbp]
\centering
\includegraphics[width=8.6cm]{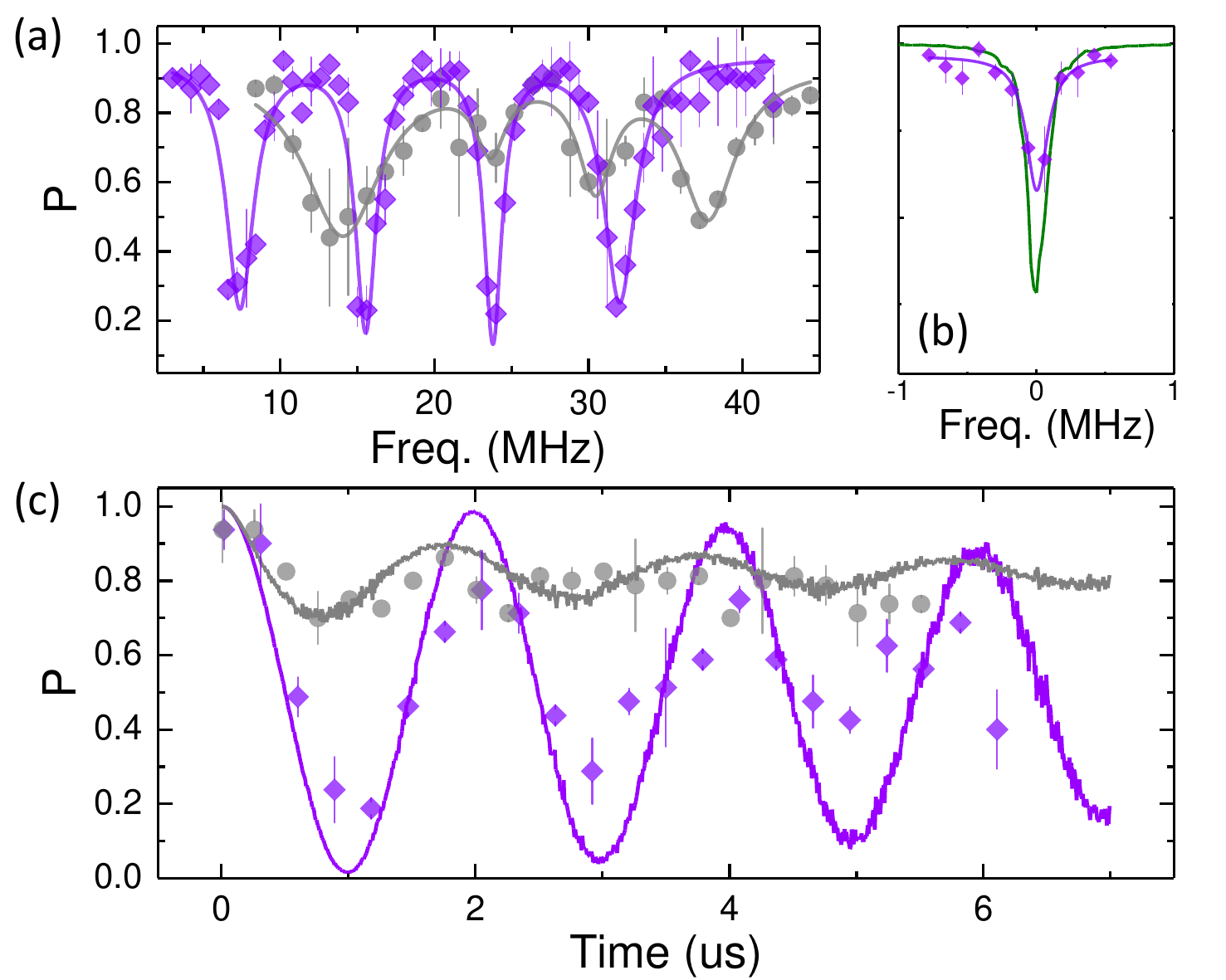}
\caption{(color online). Coherent Rydberg excitation. (a) Ground-Rydberg transition spectrum of single atoms under 4.2 Gauss bias magnetic field. The diamond (circle) points are the experimental data with (without) continuous UV light and solid curves are Lorenzian fits to the data. The four peaks in Rydberg spectra represent the ground state to 53$P_{3/2}$, $m_{j}=-3/2, -1/2, 1/2, 3/2$ state transition. The optical power of the 297-nm light $P_{297}=5.8$ mW, the pulse duration $\tau_{297}=0.2$ ms. (b) The green (purple) curve shows the calculated (measured under the UV light) spectrum. (c) Resonant Rabi oscillation between the ground-Rydberg state. The diamond (circle) points are the results with (without) the UV light. Each data points correspond to 50 repetitions of the experiment, and the error bars denote the standard deviation of the mean. The solid curves are the numerical calculation results without parameter fitting.}
\label{fig:fig2}
\end{figure}

The ground-Rydberg transition spectra of single atoms are obtained by sweeping the frequency of the 297-nm light under bias magnetic fields of 4.2 Gauss, as illustrated in Fig.~\ref{fig:fig2}(a). We compare the spectra when the 365-nm (UV) light, generated by a light-emitting diode array, is absent to the case where the vaccum cell is exposed to the UV light. Because electrons~\cite{Sedlacek2016} and a part of adsorbates~\cite{Atutov2003,Klempt2006,Telles2010,Barker2018} can be removed from the cell surface by using the UV light, these effects lead to changes of the background electric fields. We measure Stark shift $\Delta \nu=-5.4$ MHz of the $m_{j}=3/2$ peak in the presence of the UV light. The broadened linewidth of the peaks observed without the UV light is primarily due to time-dependent drifts in the background electric fields experienced by the atoms. If we attribute the broadened linewidth to the noise of the electric fields and assume it follows normal distribution, we extract the standard deviation of the distribution of the electric fields about 73 mV/cm according to our calculated results.

To estimate the residual electric field noise under the UV light, we compare our calculated linewidth with the measured value as illustrated in Fig.~\ref{fig:fig2}(b). To reduce the optical power broadening in experiment, the Rabi frequency is lowered to $2\pi\times83$ kHz. We measure the spectrum of transition $|5S_{1/2}, F=1, m_{F}=0\rangle$ $\rightarrow$ $|53P_{3/2}, m_{J}=3/2\rangle$ with a fitting linewidth of $0.19\pm0.07$ MHz. Our calculated linewidth is about 0.18 MHz under the same Rabi frequency if we consider the Doppler effect and the intensity noise except for the noise of the electric fields. The consistence indicates that the noise of the electric fields is canceled with the UV light. The stability and the reproducibility of the Rydberg spectra are described in Appendix~\ref{appendix:B}B.

We set the frequency of the 297-nm laser to the center of the peak of $|53P_{3/2}, m_{J}=3/2\rangle$ in the transition spectrum. By varying the pulse duration of the 297-nm laser, we observe Rabi oscillations with (without) UV light as the diamond (circle) points shown in Fig.~\ref{fig:fig2}(c). The coherence of the oscillation is enhanced obviously in the presence of the UV light. To analysis it, we build up models to simulate the damping of the Rabi oscillation.

The curves in Fig.~\ref{fig:fig2}(c) are the numerical calculation results if we consider the following factors (see details in Appendix~\ref{appendix:C}C). The shot-to-shot intensity noise of the 297-nm laser, the fluctuations in beam pointing and the thermal motions of the atoms all directly contribute to uncertainties in the Rabi frequency and are main decoherence factors under the UV light. Without the UV light, large frequency detunings induced by long-term drifts of the transition frequency cause uncertainties of the effective Rabi frequency and also result in severe decoherence. Other factors such as Doppler effect and the linewidth (the phase noise) of the Rydberg laser contribute less to the decay compared with those mentioned above for the Rabi oscillation, and Doppler effect can be further suppressed when the atoms are cooled into vibrational ground state~\cite{Wang2019,Lorenz2021}. Although, spontaneous emissions from the intermediate state in two-photon Rydberg experiments can be eliminated by single-photon excitation, achieving the long-coherence Rabi oscillation requires additional technical improvements, such as Raman sideband cooling and the use of low-noise lasers.

\section{electric fields induced by electrons}

\begin{figure}[htbp]
\centering
\includegraphics[width=8.6cm]{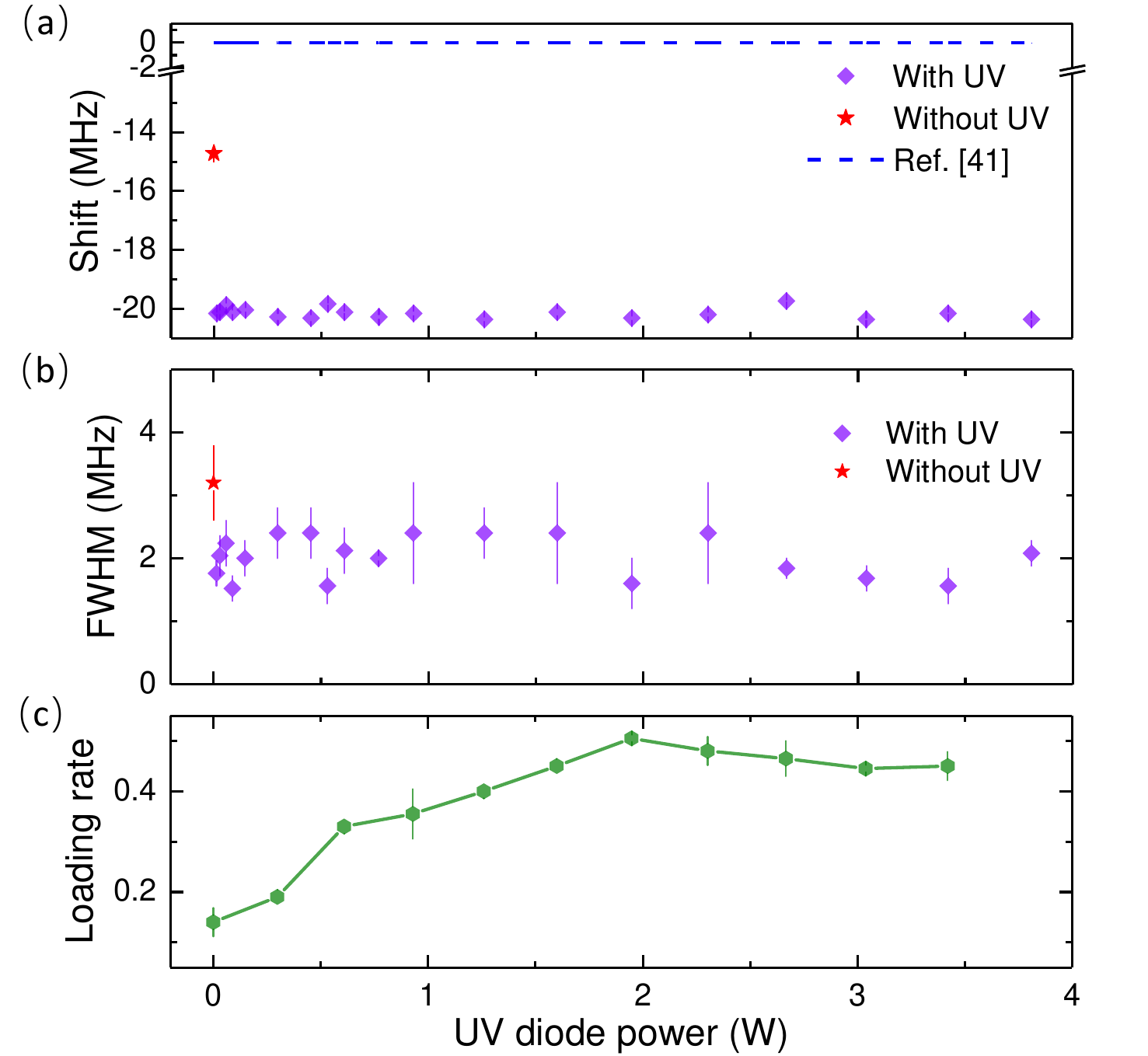}
\caption{(color online). Ground-Rydberg transition frequency with (without) the UV light. (a) The dependence of the transition frequency and the linewidth on the UV diode power, $P_{297}=5.8$ mW, $\tau_{297}=0.2$ ms. In the presence of the UV light, the vaccum cell is illuminated continuously. (b) The dependence of the linewidth on the UV diode power. (c) Average loading rates of a $6 \times 4$ tweezer array as a function of the UV diode power.}
\label{fig:fig3}
\end{figure}

To further investigate the source of the background electric fields, we explore the relation between the transition frequency of ground-Rydberg state ($|5S_{1/2}, F=1, m_{F}=0\rangle\rightarrow|53P_{3/2}, m_{J}=3/2\rangle$) and the power of the UV diodes. We observe that a power of less than 14 mW of UV light is sufficient to change the transition frequency, and no significant shift occurs as the power increases from 14 mW to 3.43 W, as illustrated in Fig.~\ref{fig:fig3}(a). A similar behavior is observed in the linewidth of the transition peaks, as depicted in Fig.~\ref{fig:fig3}(b). This result suggests that one of the primary sources of the background electric fields is the presence of electrons on the cell wall. The intensity of UV light required to remove these electrons is obviously lower than that needed for light-induced atomic desorption (LIAD)~\cite{Atutov2003,Klempt2006,Telles2010,Barker2018}. In Ref.~\cite{Sedlacek2016}, to eliminate long-lived electrons bound to the NEA surface, low-intensity UV light is used to prevent the light-induced desorption of Rb. We measure the average loading rates of the 24 tweezers~\cite{Sheng2021,Sheng2022} using various intensities of UV light, as depicted in Fig.~\ref{fig:fig3}(c). The loading rates reflect the density of background Rb atoms, which corresponds to the atoms desorbed by the UV light. To further validate this conclusion, we use 660-nm diodes to desorb the electrons without the LIAD because the bond energy of the electrons is lower than the adsorbates~\cite{Klempt2006,Sedlacek2016}. As a consequence, we observe that the loading rate remains around 0.11 when the diode power is increased, while the frequency shift and the coherent Rabi oscillation are almost the same as in the case of using the UV diode.

The Rydberg state experiences a Stark shift $\Delta \nu=-\frac{1}{2}\alpha E^{2}$ with polarizability $\alpha$=448.68 MHz/(V/cm)$^{2}$ for 53$P_{3/2}$ state, where $E$ is the value of the electric field. An experimental value $\nu=1008.731895$ THz is measured by the ground-Rydberg transition spectra under zero bias magnetic field without the UV light. The frequency of the 1188-nm light (two-step doubled to generate the 297-nm light) is precisely calibrated by the optical frequency comb technology. Dashed lines in Fig.~\ref{fig:fig3}(a) show the measured value in a vapor cell~\cite{Li2019}. If we attribute the difference between the experimental value and the absolute value of the frequency to the Stark shift induced by the electric field, and treat the value in Ref.~\cite{Li2019} as an absolute transition frequency, we extract background electric fields $E_{1}\simeq$255 mV/cm without the UV light and $E_{2}\simeq$298 mV/cm with the UV light. A reduction of the background electric field caused by the binding of electrons to the surface is consistent with the conclusion in Ref.~\cite{Sedlacek2016}.


\section{electrons generated by 297 nm light}


\begin{figure}[htbp]
\centering
\includegraphics[width=8.6cm]{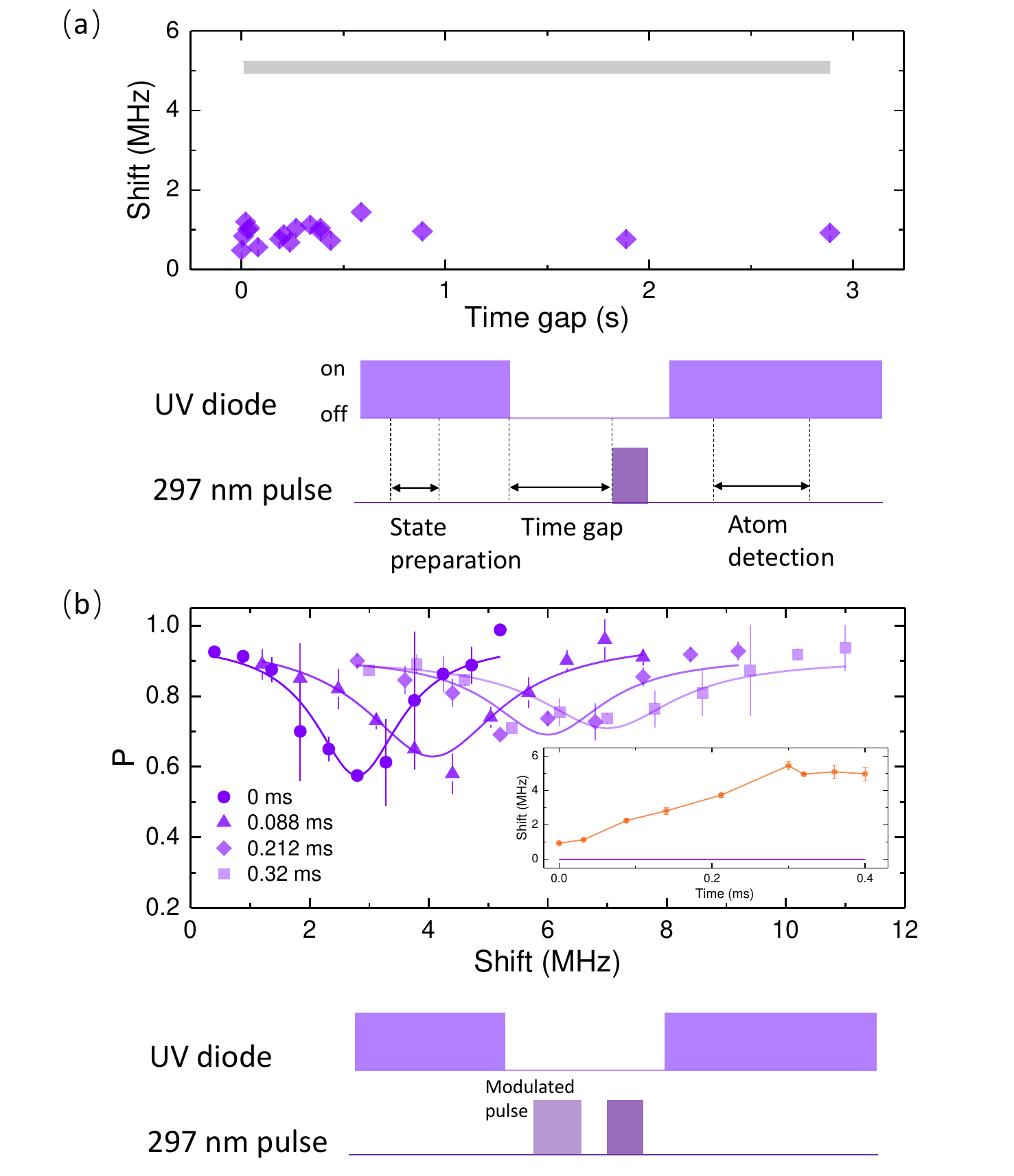}
\caption{(color online). The transition frequency dependence on electric fields induced by the 297-nm light pulse. (a) The relation between the transition frequency and the time gap. The gray line with a error bar represents the transition frequency without the UV light. The pulse diagram of the experimental sequence is in the lower panel. The pulse of the UV light is used to remove the electrons before the Rydberg excitation. The switch of the UV diode is controlled by a solid-state relay. (b) Dependence of the transition frequency on an additional 297-nm light with modulated pulse duration. $P_{297}=5.8$ mW for both two pulses, $\tau_{297}=12$ $\mu$s for the second pulse. The inset shows the relation between the transition frequency and the pulse duration. The purple line in inset represents the transition frequency with the continuous UV light. The pulse diagram of the experimental sequence is in the lower panel.}
\label{fig:fig4}
\end{figure}

To explore the source of the electrons, we first examine that whether our experimental devices for single atoms trapping can generate electrons. Fig.~\ref{fig:fig4}(a) shows that there is no obvious frequency shift after the electrons are removed by the UV light on second timescale. The unchanged electric field indicates that almost no charges are generated when the single atoms trapped in optical tweezers without the 297-nm light in our experimental setup.

Now, it appears that the electrons most likely originate from the 297-nm light after ruling out the possibility of other experimental apparatus as the source. To confirm this, we introduce an additional off-resonant 297-nm light with modulated pulse duration to specifically generate electrons. In the experimental process as the lower panel depicted in Fig.~\ref{fig:fig4}(b), the first pulse is detuned by 11.2 MHz from the frequency of the second pulse to prevent Rydberg excitation. Subsequently, the frequency of the second pulse, is scanned to measure the transition frequency. As shown in Fig.~\ref{fig:fig4}(b), the transition frequency shift increases as the duration of the first pulse is extended. This result suggests that electrons gradually accumulate on the surface when the 297-nm pulse acts on our science chamber. The mechanism behind the phenomenon of the 297-nm light generating electrons might involve blackbody ionized electrons from the Rydberg atoms~\cite{Beterov2009,Sedlacek2016,Festa2022} or partially result from Rb adsorbates with reduced ionization threshold being ionized by the 297-nm light, leading to the generation of electrons bound to the NEA surface. The precise distribution of the electrons and a comprehensive understanding of the electron photodesorption process on the NEA surface require further investigation in the future.

After a source of the background electric fields is revealed, we interpret the noise of the electric fields is probably induced by the unstable existence of the electrons on the cell surface. Another type of the frequency detunings caused by the position instability of the single atoms is negligible in our case, because the distance between single atoms and the cell wall is approximately three orders of magnitude larger than the position fluctuations of single atoms in our experiment. But it becomes one of the major decoherence factors when Rydberg atoms approach a charged surface~\cite{Ocola2022}. Additionally, we should note that applying a stray electric field by electrodes for compensating the total field to zero is another method to enhance coherence~\cite{Kubler2010,Graham2022}. Since $\delta\Delta\nu(E)=-\alpha E\delta E$, the transition frequency becomes less sensitive to variations in the electric field $E$ as the electric field decreases. However, this method may not be as effective as eliminating the noise source through electron photodesorption, particularly when an additional electric field is required for electrically-tuned F\"{o}rster resonance.


\section{Collective oscillations}

\begin{figure}[htbp]
\centering
\includegraphics[width=8.6cm]{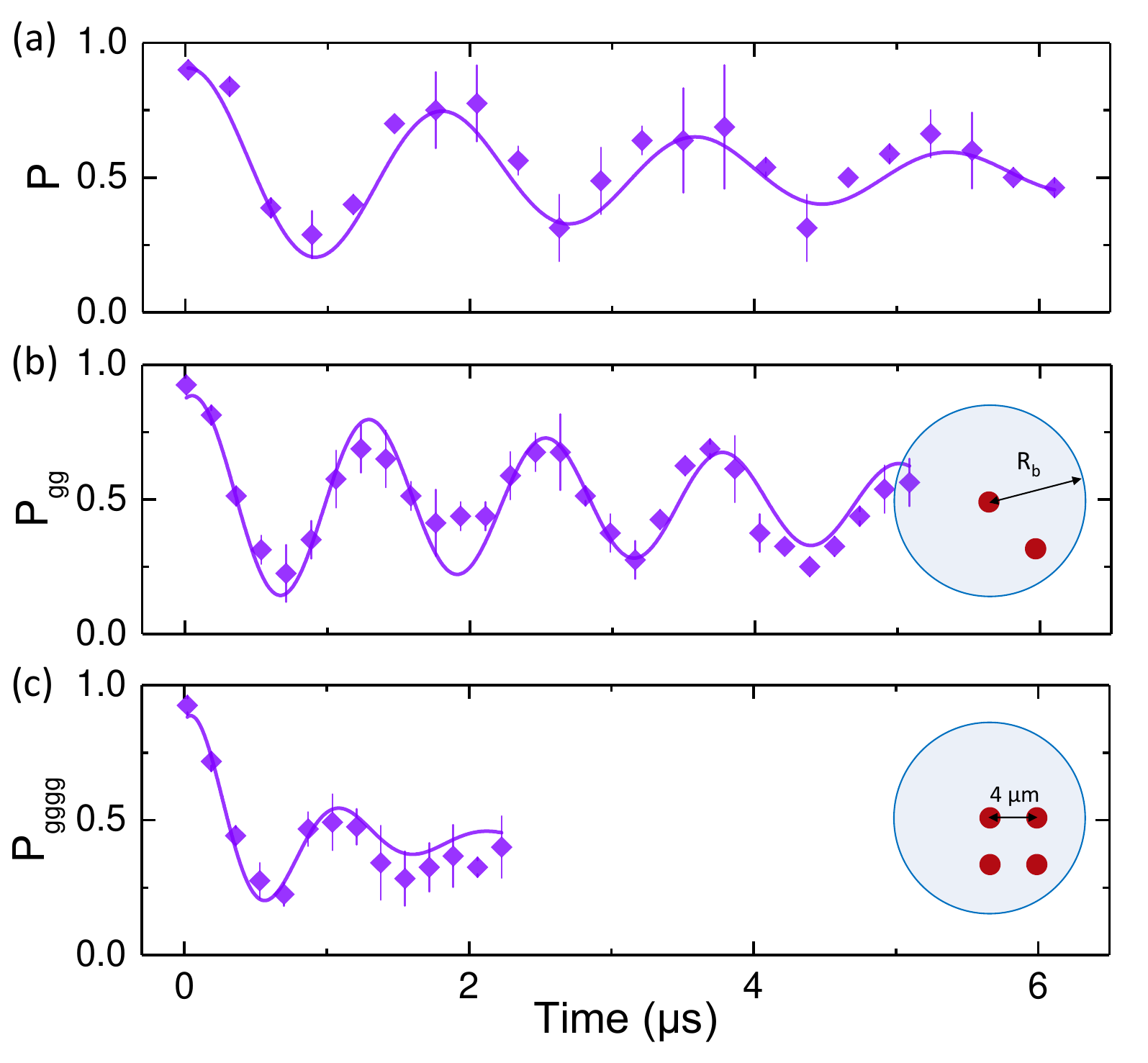}
\caption{(color online). Collective oscillations. (a) Resonant Rabi oscillation between the ground-Rydberg state of a single atom. The solid purple curve is a damped sinusoidal fit of the oscillation, yielding the Rabi frequencies of 2$\pi \times$ 0.71 MHz. (b) Collective oscillations of two atoms. $P_{gg}$ represents the probability of the two atoms both in ground state. (c) Collective oscillations of four atoms.}
\label{fig:fig5}
\end{figure}

Furthermore, we demonstrate collective oscillations in the full Rydberg blockade regime of two (four) atoms as illustrated in Fig.~\ref{fig:fig5}(b) (Fig.~\ref{fig:fig5}(c)), respectively. The
blockade radiu $R_{b}=7.5$ $\mu$m and the distance between two neighboring sites is about 4 $\mu$m. In this experiment, the atoms are determined prepared in optical tweezers via atom rearrangement and initialized in $|5S_{1/2}, F=1, m_{F}=0\rangle$. The atoms are then driven by global Rydberg excitation pulses with balanced Rabi frequency. We measured the probability that all atoms are in ground state as a function of the pulse duration. A frequency enhanced by a factor $\sim\sqrt{N}$ with respect to the single-atom case (Fig.~\ref{fig:fig5}(a)) is observed.


\section{conclusion}


In summary, optically trapped single atoms are used as in-situ sensors to measure the background electric fields. We reveal that the electrons influencing the background electric fields are generated by the 297-nm light. These accumulated electrons lead to frequency shifts and broadened transition peaks during Rydberg excitation. After they are removed from the cell surface through exposure to the UV light, the time-dependent electric field noise is eliminated by electron photodesorption process. Consequently, the coherence of the ground-Rydberg transition is enhanced. Our investigations will advance the control and reliability of Rydberg atom-based systems, allowing them to be applied across a wide range of quantum technologies, such as high-fidelity multi-qubit gates, quantum simulators based on Rydberg dressing, and precise quantum sensors.


\section{acknowledgments}


This work was supported by the National Natural Science Foundation of China under grants No. 12122412, No. U22A20257, and No. 12121004, No. 12004395, No. 12104464, No. 12241410, the Project for Young Scientists in Basic Research of CAS under grant No. YSBR-058, the Key Research Program of Frontier Science of CAS under grant No. ZDBSLY-SLH012, the National Key Research and Development Program of China under Grant No. 2017YFA0304501, and the Youth Innovation Promotion Association of CAS under grant No. 2019325. The authors thank Q.-F. Chen for technique support of the optical frequency comb.

B. M. and C. S. contributed equally to this work.


\section{appendix a: 297-nm laser system}\label{appendix:A}

\begin{figure*}[htbp]
\centering
\includegraphics[width=16cm]{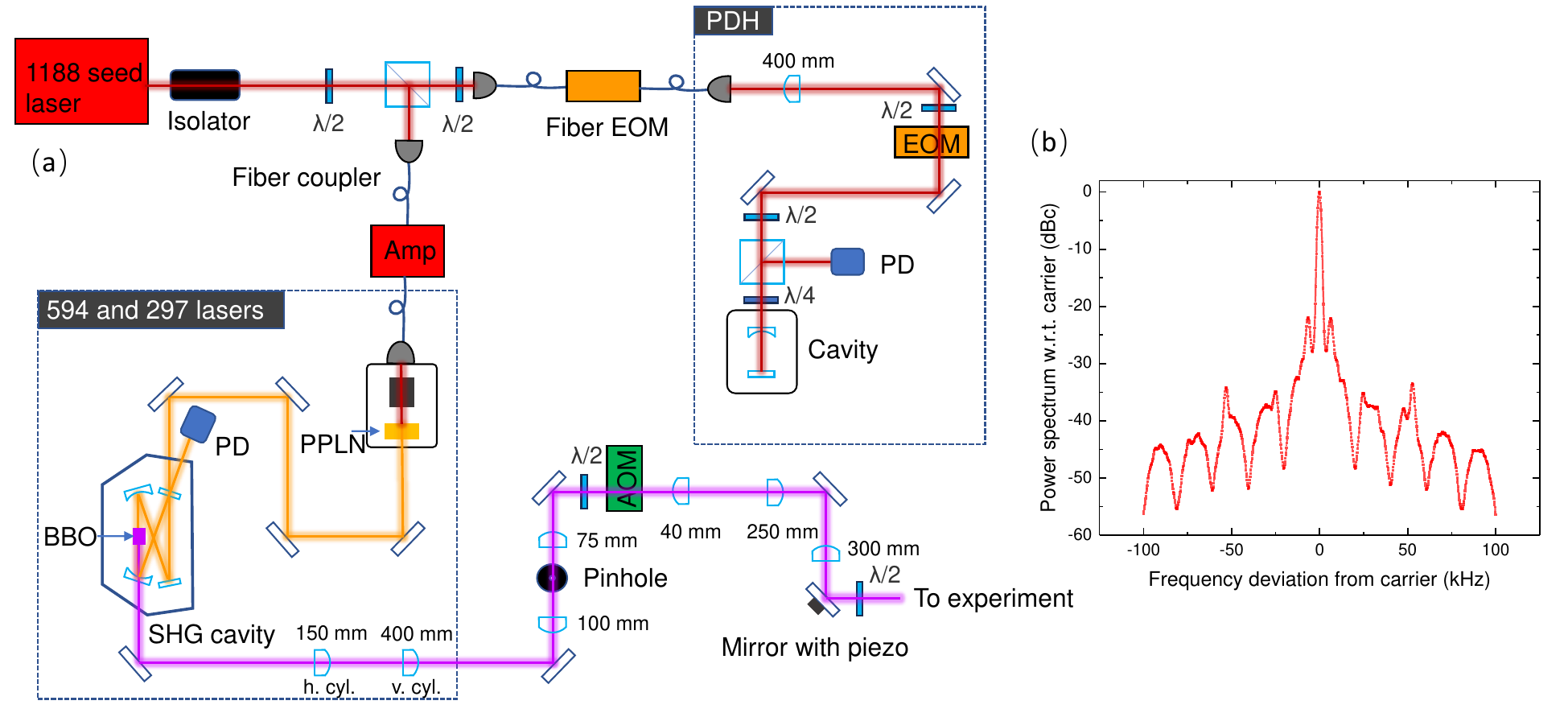}
\caption{(color online). (a) A schematic diagram of the 297-nm laser system. (b) Measured self-heterodyne spectrum of 1188-nm laser with a 10-km fiber delay.}
\label{fig:fig6}
\end{figure*}

In this section, we provide detailed information about our 297-nm laser system for Rydberg excitation. This system is constructed through a two-step frequency doubling process, as illustrated in Fig.~\ref{fig:fig6}(a). The 1188-nm light, generated by a seed laser (Toptica DL pro), is first amplified using a Raman amplifier (PRECILASERS RFL-FHG 297-CW). Next, it is directed into a periodically poled lithium niobate (PPLN) crystal ($25\times7.9\times0.5$ mm) at a temperature of $49.82^\circ$C. The 1188-nm light is then converted into 594-nm light with an optical power of 1.4 W, using a 9.8 A current for the amplifier. A small fraction of this 594-nm light, totaling a few milliwatts, is directed to a wavemeter (WS7-60) to monitor the laser frequency for Rydberg excitation. In the second harmonic generation stage, the optical frequency is doubled from 594-nm to 297-nm light. This is achieved by passing the 594-nm light through a beta barium borate (BBO) crystal ($4\times4\times10$ mm) in a second-harmonic generation (SHG) cavity.

Firstly, we employ a pair of cylindrical mirrors to reshape the 297-nm beam along both the horizontal and vertical axes. Secondly, the initial 297-nm beam is focused and directed through a pinhole with a diameter of 50 $\mu$m. This step is essential for removing transverse structures in the laser mode, and it yields a well-suited Gaussian beam through spatial filtering. The polarization of the beam prior to entering the AOM is linear and can be adjusted using a Glan-Taylor polarizer and a half-wave plate. Before focusing the UV light on the atom, it is expanded using 40 mm and 250 mm plano-convex lenses. The expanded beam is subsequently focused into an elliptical shape, with $1/e^{2}$ radius of 14.2 (22.5) $\mu$m in horizontal (vertical) axes. This is achieved using a 300 mm plano-convex lens. The direction of the beam is precisely controlled using a mirror equipped with a piezo.

Frequency stabilization is critical for Rydberg experiments. To achieve this, we utilize the Pound-Drever-Hall (PDH) technique, which allows us to obtain a laser with a narrow linewidth suitable for Rydberg excitation. The measured finesse of the Fabry-Perot cavity is $5462\pm3$. We employ a home-built electro-optic modulator (EOM) to generate 8.3 MHz sidebands for frequency modulation. These sidebands interact with the cavity, resulting in reflections that are detected by a photodetector (PD). The PD signal is demodulated to produce a PDH error signal, which is used to lock the laser frequency.To scan the frequency of the 297-nm light, we employ a fiber EOM (PM-0S5-10-PFA-PFA-1188) to generate $\pm1$ sidebands on the 1188-nm light. By locking the frequency of the +1 sideband and modulating the driven frequency of the fiber EOM, we can tune the frequency of the 297-nm light within a range of $\pm200$ MHz. We measure the self-heterodyne spectrum of the 1188-nm laser using a 10-km fiber delay line when the locked loop is active. The acquired linewidth is 563.1 Hz, as illustrated in Fig.~\ref{fig:fig6}(b). This demonstrates that we can achieve 297-nm light with a linewidth of 2.2 kHz, which is suitable for Rydberg excitation.


\section{appendix b: stability and reproducibility of Rydberg spectra}\label{appendix:B}


We observe that the Rydberg spectra are reproducible and remain stable during several months with the UV light. The spectral shift is less than 2.8 MHz in almost one year even if we try to heat the vacuum cell by hot wind to change the distribution of adsorbates or remove charges from the outer surface of the cell wall. However, the spectral shift can reach about 6 MHz without the UV light and occasionally changes during one week. The variation rate of the electric fields as shown in Fig.~\ref{fig:fig3}(b) is unstable, ranging from several hundreds of microseconds to several tens of milliseconds. It seems to be relevant to the background Rb atoms in the cell, because the rate always undergoes changes after we turn on the Rb dispenser for several hours (the dispenser is about 32 cm far from the center of the vacuum cell and is turned on at the current of 2.6 A).

Additionally, we also attempt to use 465-nm diode for electron photodesorption, and almost the same behaviors are observed as with the 365-nm diode.


\section{appendix c: origins of decoherence}\label{appendix:C}


\begin{figure}[htbp]
\centering
\includegraphics[width=8.6cm]{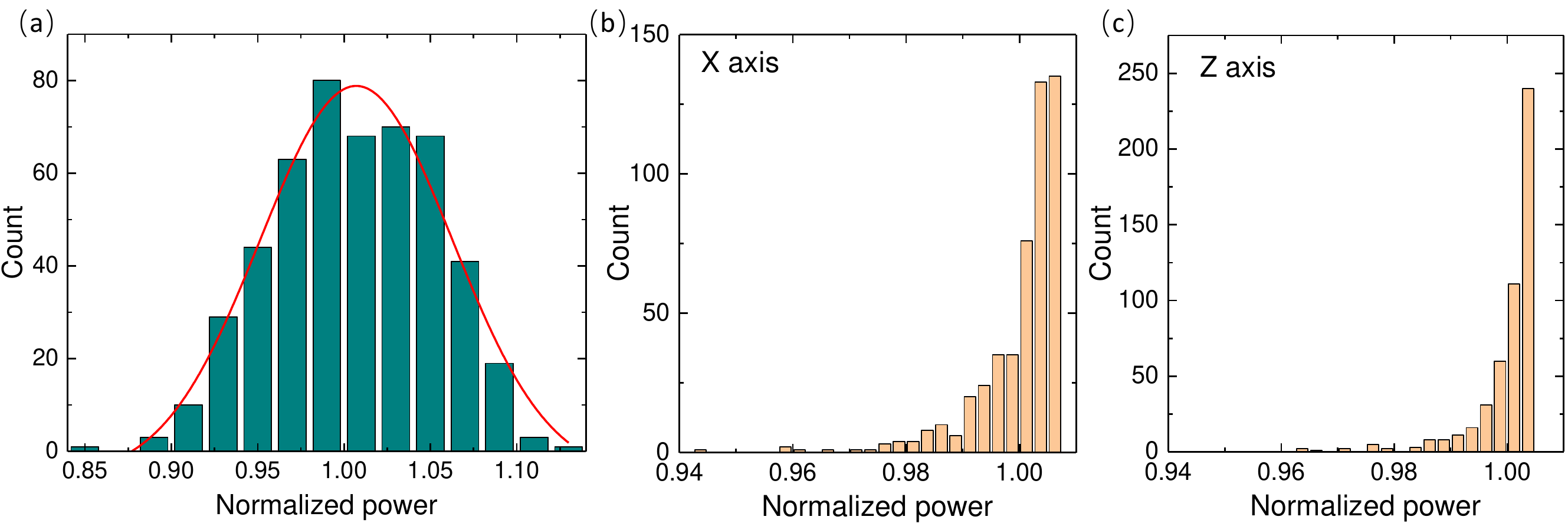}
\caption{(color online). (a) Shot-to-shot intensity noise of the 297-nm laser. (b) Pointing stability in X axis of the focused 297-nm laser beam. (c) Pointing stability in Z axis.}
\label{fig:fig7}
\end{figure}

Candidates for decoherence mechanisms of Rydberg-ground Rabi oscillation include intensity noise of the 297-nm laser (1), fluctuations in beam pointing of the 297-nm laser (2), thermal motions of the atoms (3), frequency detunings induced by long-term drifts of the transition frequency without the UV light (4) and Doppler effect (5). We calculate the Rabi oscillation by a model defined as

\begin{equation}\label{eq1}
P(t)=\frac{\Omega^{2}}{\Omega^{2}+\Delta^{2}}\sin^{2}(\sqrt{\Omega^{2}+\Delta^{2}}\frac{t}{2}),
\end{equation}
where $P(t)$ denotes the probability of the atoms in Rydberg state at time $t$. $\Omega$ and $\Delta$ represent the Rabi frequency and frequency detuning, respectively.

(1) $\textit{Intensity noise of the 297-nm laser}$. The acquired distribution of the shot-to-shot intensity noise obeys normal distribution with $\sigma_{I}/I_{0}=0.06$ (Fig.~\ref{fig:fig7}(a)), where $\sigma_{I}$ denotes a standard deviation of the distribution and $I_{0}$ denotes the normalized power. This intensity noise is mainly resulted from the amplifying process of the 1188-nm light.

(2) $\textit{Fluctuations in beam pointing of the 297-nm light}$. The pointing fluctuations of 297-nm light are 1.1 (1.4) $\mu$m in X (Z) axes measured by Dual Scanning Slit Beam Profilers (BP209-VIS/M) in 101 seconds. The intensity distribution experienced by atoms induced by the pointing instability are shown in Fig.~\ref{fig:fig7}(b) ((Fig.~\ref{fig:fig7}(c)) in X (Z) axes.

(3) $\textit{Thermal motions of the atoms}$. In our experiment, the motion of the atoms results in position fluctuations with calculated uncertainties of 1.3 (0.2) $\mu$m in X (Z) axes.

(4) $\textit{Drifts of the transition frequency}$. Without the UV light, the uncertainties of transition frequency is evaluated by the numerical simulation results of the broadened linewidth. If we assume them follow normal distribution, we obtain a standard deviation of 8.2 MHz.

(5) $\textit{Doppler effect}$. Doppler shift $\Delta_{D}$ of the atoms is given by $\Delta_{D}= k\cdot v$, where the wave vector $k = 2\pi/\lambda$. $\lambda$ is the wavelength of the Rydberg excitation laser and the $v$ is the atom velocity in direction of the 297-nm laser beam which is given by $v=\sqrt{k_{B} T/m}$ with the atom temperature $T=3.5$ $\mu$K and the mass $m$ of the atoms. We calculate $\Delta_{D}=2\pi\times$ 61.5 kHz.

For the effective Rabi frequency $\Omega_{eff}=\sqrt{\Omega^{2}+\Delta^{2}}$ in Eq. (1), decoherence factors (1)(2)(3) contribute to the uncertainties of $\Omega$, and (4)(5) contribute to the uncertainties of $\Delta$. Considering these factors, we obtain the calculated Rabi oscillation as shown in Fig.~\ref{fig:fig2}(b). Because long-term drifts (on a time scale of several hours) of the pointing of the 297-nm light cause a misalignment, $\Omega$ becomes more sensitive to the pointing fluctuations when the atoms are located away from the center of the 297-nm laser beam.  This would be the main reason for the measured Rabi oscillation with a shorter coherence time than our calculated results.



\end{document}